\documentclass[useAMS,usenatbib]{mn2e}

\usepackage{graphicx}

\newcommand       \be           {\begin{equation}}
\newcommand       \ee           {\end{equation}}
\newcommand       \bea          {\begin{eqnarray}}
\newcommand       \eea          {\end{eqnarray}}
\newcommand       \apj          {ApJ}
\newcommand       \apjl         {ApJL}
\newcommand       \aap          {A\&A}
\newcommand       \nat          {Nature}
\newcommand       \mnras        {MNRAS}
\def\simlt{\mathrel{\hbox{\rlap{\hbox{\lower4pt\hbox{$\sim$}}}\hbox{$<$}}}}
\def\simgt{\mathrel{\hbox{\rlap{\hbox{\lower4pt\hbox{$\sim$}}}\hbox{$>$}}}}
\def\lesssim{\mathrel{\hbox{\rlap{\hbox{\lower4pt\hbox{$\sim$}}}\hbox{$<$}}}}

\def\gtrsim{\mathrel{\hbox{\rlap{\hbox{\lower4pt\hbox{$\sim$}}}\hbox{$>$}}}}

\title[Nickel-Rich AIC Outflows]{Nickel-rich outflows from accretion disks formed by the accretion-induced collapse of white dwarfs} \author[B.~D. Metzger, A.~L. Piro, E. Quataert]{B.~D. Metzger\thanks{E-mail:
bmetzger@astro.berkeley.edu}, A.~L. Piro, and E. Quataert \\
Astronomy Department and Theoretical Astrophysics Center,
University of California, Berkeley, 601 Campbell Hall, Berkeley CA,
94720\\}
\begin{document}
\date{Accepted . Received ; in original form }
\pagerange{\pageref{firstpage}--\pageref{lastpage}} \pubyear{????}
\maketitle
\label{firstpage}

\begin{abstract}

A white dwarf (WD) approaching the Chandrasekhar mass may in several circumstances undergo accretion-induced collapse (AIC) to a neutron star (NS) before a thermonuclear explosion ensues.  It has generally been assumed that such an accretion-induced collapse (AIC) does not produce a detectable supernova (SN).  If, however, the progenitor WD is rapidly rotating (as may be expected due to its prior accretion), a centrifugally-supported disk forms around the NS upon collapse.  We calculate the subsequent evolution of this accretion disk and its nuclear composition using time-dependent height-integrated simulations with initial conditions taken from the AIC calculations of Dessart et al.~(2006).  Soon after its formation, the disk is cooled by neutrinos and its composition is driven neutron-rich (electron fraction $Y_{e} \sim 0.1$) by electron captures.  However, as the disk viscously spreads, it is irradiated by neutrinos from the central proto-NS, which dramatically alters its neutron-to-proton ratio.  We find that electron neutrino captures increase $Y_{e}$ to $\sim 0.5$ by the time that weak interactions in the disk freeze out.  Because the disk becomes radiatively inefficient and begins forming $\alpha$-particles soon after freeze out, powerful winds blow away most of the disk's remaining mass.  These $Y_{e} \sim 0.5$ outflows synthesize up to a few times $10^{-2}M_{\sun}$ in $^{56}$Ni.  As a result, AIC may be accompanied by a radioactively-powered SN-like transient that peaks on a timescale of $\sim 1$ day.  Since few intermediate mass elements are likely synthesized, these Nickel-rich explosions should be spectroscopically distinct from other SNe.  The  timescale, velocity, and composition of the AIC transient can be modified if the disk wind sweeps up a $\sim 0.1 M_{\sun}$ remnant disk created by a WD-WD merger; such an ``enshrouded'' AIC may account for sub-luminous, sub-Chandrasekhar Type I SNe.  Optical transient surveys such as PanSTARRS and the Palomar Transient Factory should detect a few AIC transients per year if their true rate is $\sim 10^{-2}$ of the Type Ia rate, and the Large Synoptic Survey Telescope (LSST) should detect several hundred per year.  High cadence observations ($\lesssim 1$ day) are optimal for the detection and follow-up of AIC.
\end{abstract}


\vspace{-3cm}
\section{Introduction}
\voffset=-2cm
\vspace{-0.2 cm}
\label{sec:int}
\label{sec:intro}

It is generally thought that when an accreting C-O white dwarf (WD) approaches the Chandrasekhar mass $M_{\rm Ch}$, the carbon ignites in its core, causing the WD to explode and produce a Type Ia supernova (SN).  If, however, the WD's composition is O-Ne instead of C-O, the outcome may be qualitatively different: electron captures on nuclei will cause the WD to collapse to a neutron star (NS) before an explosion can ensue (Miyaji et al.~1980; Canal et al.~1990; Gutierrez et al.~2005; Poelarends et al.~2008).  Such an accretion-induced collapse (AIC) may also occur for C-O WDs with large initial masses ($M > 1.1M_{\sun}$) in systems with mass-transfer rates $\dot{M} \gtrsim 10^{-8}M_{\sun}$ yr$^{-1}$ (Nomoto $\&$ Kondo 1991) or following the merger of two WDs in a binary if their total mass exceeds $M_{\rm ch}$ (e.g., Mochkovitch $\&$ Livio 1989, 1990; Yoon et al.~2007).  Indeed, spectroscopic surveys of WD-WD binaries (e.g., SPY; Napiwotzki et al.~2004) have discovered at least one system that will merge within a few Hubble times with a total mass tantalizingly close to $M_{\rm Ch}$ (Napiwotzki et al.~2002).

Despite its likely occurrence in nature, AIC has not yet been observationally identified, probably because it is less common and produces less $^{56}$Ni than a normal SN (e.g., Woosley $\&$ Baron 1992).  Identifying AIC or constraining its rate would, however, provide unique insights into the paths of degenerate binary evolution, the true progenitors of Type Ia SNe (e.g., Yungelson $\&$ Livio 1998), and the formation channels of globular cluster NSs (Grindlay 1987), low mass X-ray binaries (van den Heuvel 1984; Michel 1987), and millisecond pulsars (Grindlay $\&$ Bailyn 1988; Bailyn $\&$ Grindlay 1990).  Identifying the optical signature of AIC is also important because these events may be a source of strong gravitational wave emission (Fryer et al.~2002; see Ott 2008 and references therein).

In this Letter we show how AIC may itself produce a SN-like transient which, although less luminous and shorter-lived than a typical SN, may nonetheless be detectable with upcoming optical transient surveys.  Although little $^{56}$Ni is ejected by the newly-formed proto-NS itself, we show that powerful winds from the accretion disks formed during AIC produce up to a few times $10^{-2}M_{\sun}$ in $^{56}$Ni.  

\vspace{-0.6 cm}
\subsection{Nickel-Rich Winds from AIC Disks}
\label{sec:bigpicture}
\vspace{-0.2 cm}
\begin{figure}
\resizebox{\hsize}{!}{\includegraphics{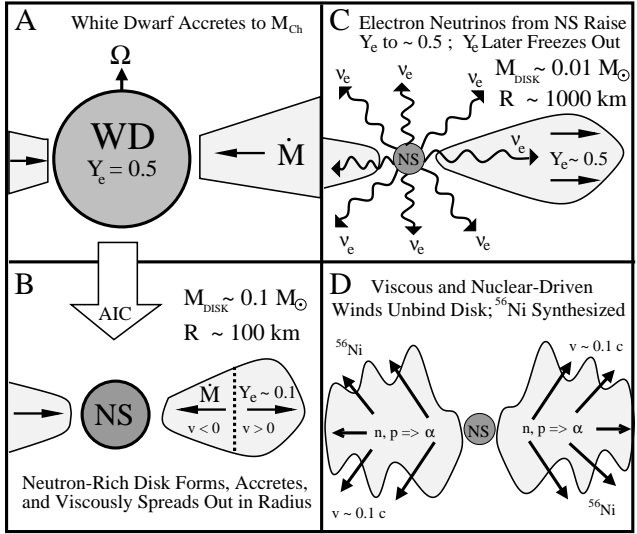}}
\caption{Stages in the accretion-induced collapse of a WD.  (A) WD accretes up to near the Chandrasekhar mass ($M_{\rm Ch}$) from either a binary companion or following the disruption of another WD in a merger.  Due to the accretion of angular momentum, the WD is rapidly rotating prior to collapse. (B) Electron captures cause the WD to collapse to a proto-NS surrounded by a compact accretion disk with a mass $\sim 0.1 M_{\sun}$ and a size $\sim 100$ km.  Electron captures drive the composition neutron-rich (i.e., to an electron fraction $Y_{e} \sim 0.1$).  As matter begins to accrete onto the proto-NS, the bulk of the disk spreads to larger radii.
(C)  As the disk spreads, the midplane is irradiated by neutrinos from the central proto-NS.  Electron neutrinos are absorbed by neutrons, raising the electron fraction to $Y_{e} \sim 0.5$.  (D) The disk's electron fraction freezes out.  Soon thereafter, powerful outflows blow the disk apart.  As the ejecta expands adiabatically, material with $Y_{e} \gtrsim 0.5$ is efficiently synthesized into $^{56}$Ni, the subsequent decay of which powers a SN-like optical transient.}
\label{fig:cartoon}
\end{figure}

Figure $\ref{fig:cartoon}$ summarizes the stages of AIC.  Due to their preceding accretion, WDs that undergo AIC are probably rotating rapidly prior to collapse (Fig.~$\ref{fig:cartoon}$A).  As a result, a substantial fraction of the WD mass must be ejected into a disk during the collapse in order to conserve total angular momentum (Shapiro $\&$ Lightman 1976; Michel 1987).  Dessart et al.~(2006, 2007; hereafter D06,07) perform 2D axisymmetric AIC calculations, in which they find that a quasi-Keplerian accretion disk with mass $M_{\rm d,0} \sim 0.1-0.5M_{\sun}$ forms around the newly-formed proto-NS.  

In Metzger, Piro, $\&$ Quataert (2008a,b; hereafter MPQ08a,b) we studied the time-dependent evolution of accretion disks with similar initial properties that are formed from black hole (BH)-NS or NS-NS mergers.  Many of these results also apply to proto-NS accretion following AIC.  Although the disk is initially concentrated just outside the proto-NS surface, the bulk of the disk's mass must spread to larger radii as matter accretes in order to conserve angular momentum.  Early in their evolution, such ``hyper-accreting'' disks are neutrino-cooled and geometrically thin, and their composition is driven neutron-rich by an equilibrium between electron and positron captures under degenerate conditions (e.g., Pruet et al.~2003; Fig.~$\ref{fig:cartoon}$B): the electron fraction $Y_{e} \sim 0.1$, where  $Y_{e} \equiv n_{p}/(n_{p}+n_{n})$ and $n_{p}/n_{n}$ are the proton/neutron number densities.  As the disk viscously spreads and its temperature decreases, however, weak interactions eventually become slow compared to the evolution timescale of the disk and the value of $Y_{e}$ freezes out.  

Soon following freeze-out, neutrino cooling becomes inefficient and the disk becomes advective and geometrically thick (MPQ08a,b).  Since advective disks are only marginally bound, powerful winds likely begin to unbind most of the disk's remaining mass (Blandford $\&$ Begelman 1999), aided by the nuclear energy released as $\alpha-$particles begin to form (Lee $\&$ Ramirez-Ruiz 2007; MPQ08a,b).  Because $\sim 20\%$ of the initial disk mass $M_{d,0}$ still remains when the disk becomes advective (or when $\alpha-$particles form), a robust consequence of disk formation during AIC is the ejection of up to a few times $10^{-2}M_{\sun}$.  Since the ejecta are hot and dense, heavy isotopes are synthesized as they expand away from the midplane and cool.  Because weak interactions are already slow by this point, which heavy isotopes are produced depends on $Y_{e}$ in the disk at freeze out. 

In the case of BH accretion following BH-NS or NS-NS mergers, the disk freezes out neutron-rich with an electron fraction $Y_{e} \sim 0.2-0.4$ (MPQ08a,b).  A crucial difference in the case of AIC, however, is the presence of the central proto-NS, which radiates a substantial flux of electron neutrinos as it deleptonizes during the first few seconds following core bounce (e.g., Burrows $\&$ Lattimer 1986).  This $\nu_{e}$ flux irradiates the disk, which acts to raise $Y_{e}$ via absorptions on free neutrons ($\nu_{e} + n \rightarrow e^{-} + p$; Fig.~\ref{fig:cartoon}C).

In $\S\ref{sec:disk_evo}$ we present calculations of the evolution of the accretion disks formed from AIC using time-dependent height-integrated simulations with initial conditions and neutrino irradiation taken from the AIC simulations of D06.  We show that due to $\nu_{e}$ absorptions, a substantial fraction of the disk freezes out with $Y_{e} \gtrsim 0.5$.  As a result, late time outflows from the disk primarily synthesize $^{56}$Ni (Fig.~\ref{fig:cartoon}D).  We discuss the observational signature of these Ni-rich outflows in $\S\ref{sec:miniSN}$, and we conclude in $\S\ref{sec:rates}$ by evaluating the prospects for detecting AIC with upcoming optical transient surveys and as electromagnetic counterparts to gravitational wave sources.
\vspace{-1 cm}
\section{AIC Accretion Disk Model}
\label{sec:disk_evo}
\vspace{-0.2 cm}
\subsection{Methodology}
\vspace{-0.2 cm}
The equations and assumptions employed here closely follow those in $\S$3.1 of MPQ08b for NS-NS or BH-NS mergers.  To provide a brief summary, we evolve the surface density $\Sigma$, temperature $T$, and electron fraction $Y_{e}$ as a function of radius $r$ and time $t$ using the 2N-RK3 scheme described in Brandenburg (2001).  We employ an ``$\alpha$'' prescription for the viscosity (Shakura $\&$ Sunyaev 1973), with the accretion stress proportional to the gas pressure.  Our calculation includes all of the relevant neutrino cooling processes for both optically thin and optically thick disks (e.g., Di Matteo et al.~2002).  We take the initial surface density $\Sigma(r,t=0)$ from the Keplerian disk that forms soon after core bounce in the AIC calculations of D06.  We focus on the 1.46$M_{\sun}$ WD model of D06, which collapses to form a $\approx 0.1 M_{\sun}$ disk that extends from the NS surface at $\approx 30$ km out to several NS radii (see Fig.~5 of D06).

We evolve $Y_{e}$ including both local $e^{-}/e^{+}$ pair captures ($e^{-}+p \rightarrow \nu_{e}+n$ and $e^{+}+n \rightarrow \bar{\nu}_{e}+p$) and non-local neutrino absorptions ($\nu_{e} + n \rightarrow e^{-} + p$ and $\bar{\nu}_{e} + p \rightarrow e^{+} + n$) due to irradiation from the central proto-NS.  We approximate the neutrino flux from the proto-NS as arising from a ``light bulb'' at the origin because the neutrino capture rates only become important relative to the pair capture rates when the disk has spread to a radius which is much larger than that of the central NS.  We neglect the disk's neutrino luminosity because it is small compared to the proto-NS luminosity.  We assume that the disk's initial electron fraction is $Y_{e}(r,t=0) = 0.5$, i.e., that of the WD prior to collapse.

The neutrino absorption rates are proportional to $L_{\nu}\langle\epsilon_{\nu}\rangle$, where $L_{\nu}$ and $\langle\epsilon_{\nu}\rangle$ are the neutrino luminosities and mean energies, respectively, from the proto-NS.  During the first $t \approx 1$ s following core bounce, we use $L_{\nu_{e}},L_{\bar{\nu}_{e}},\langle\epsilon_{\nu_{e}}\rangle,$ and $\langle\epsilon_{\bar{\nu}_{e}}\rangle$ from D06 (specifically, their values in the equator; see their Fig.~9).  The $L_{\nu_{e}}$ flux peaks at a very large value ($\gtrsim 10^{53}$ ergs s$^{-1}$) at early times, after the shock created at core bounce ``breaks out'' of the proto-NS's neutrinosphere.  The flux at later times $(t \gtrsim 0.1$ s) is dominated by the cooling proto-NS, with $L_{\nu_{e}}$ remaining $\gtrsim L_{\bar{\nu}_{e}}$ as the proto-NS deleptonizes (Burrows $\&$ Lattimer 1986).  Note that because $L_{\nu_{e}}\langle\epsilon_{\nu_{e}}\rangle > L_{\bar{\nu}_{e}}\langle\epsilon_{\bar{\nu}_{e}}\rangle$ during most of the first second following core bounce, the rate of $\nu_{e}$ captures exceeds the rate of $\bar{\nu}_{e}$ captures and the net effect of neutrino absorptions is to drive $Y_{e}$ to a value $\gtrsim 0.5$ (Qian et al.~1993).  This result is robust since shock break-out and deleptonization are generic features of NS formation (e.g., Burrows $\&$ Mazurek 1983); indeed, we find similar results using luminosities and mean energies from the rotating core-collapse simulations of Thompson, Quataert, $\&$ Burrows (2005).  

When necessary, for $t\gtrsim 1$ s we take $L_{\nu} \propto t^{-1}$, motivated by the calculations of Pons et al.~(1999), and assume $\langle\epsilon_{\nu}\rangle \propto L_{\nu}^{1/4} \propto t^{-1/4}$, as appropriate for a blackbody with a fixed radius.  Although the neutrino luminosities and spectra at such late times are somewhat uncertain, we find that for $\alpha \gtrsim 0.01$ the disk evolves sufficiently rapidly that our results are relatively insensitive to the precise values of $L_{\nu}$ and $\langle\epsilon_{\nu}\rangle$ at late times.     

\vspace{-0.7 cm}
\subsection{Results}
\vspace{-0.2 cm}
\begin{figure}
\resizebox{\hsize}{!}{\includegraphics{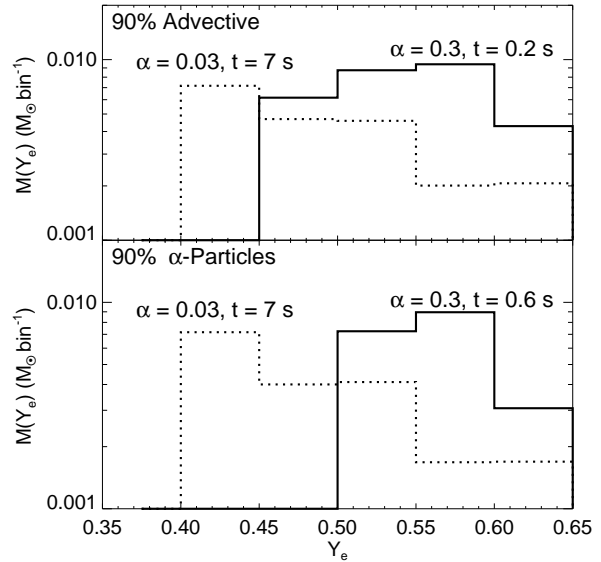}}
\caption{Amount of mass with a given electron fraction $M(Y_{e})$ in the accretion disk formed from AIC for a model with an initial mass distribution and neutrino irradiation from the $1.46 M_{\sun}$ WD AIC simulation of D06.  The top and bottom panels correspond to times when $90\%$ of the disk's mass is advective and has formed $\alpha-$particles, respectively.  The solid and dashed lines show results for $\alpha = 0.3$ and $\alpha = 0.03$, respectively.}
\label{fig:yef}
\end{figure}

The viscous and thermal evolution of the accretion disks formed from AIC are very similar to those presented in MPQ08b for NS-NS/NS-BH mergers, so we refer the reader to this work for a detailed discussion of the overall evolution of the disk.  The major qualitative difference between AIC and NS-NS/NS-BH merger disks is the evolution of the electron fraction, so we focus on our results for $Y_{e}$ below.

Figure $\ref{fig:yef}$ shows histograms of mass as a function of electron fraction $M(Y_{e})$ for our fiducial model that employs an initial density profile and neutrino irradiation from the 1.46$M_{\sun}$ WD model of D06.  The solid and dashed lines show calculations performed with viscosity $\alpha = 0.3$ and $\alpha = 0.03$, respectively.  The top panel shows $M(Y_{e})$ when $90\%$ of the disk's mass has become advective, which corresponds to a time $t = 0.2(7)$ s for $\alpha = 0.3(0.03)$.  Annuli in the disk are considered advective once the neutrino cooling rate becomes $\lesssim 1/2$ of the viscous heating rate.  The bottom panel shows $M(Y_{e})$ when $90\%$ of the disk's mass has fused into $\alpha-$particles, which corresponds to a time $t = (0.6)7$ s for $\alpha = 0.3(0.03)$.  Although powerful outflows likely begin when the disk becomes advective (Blandford $\&$ Begelman 1999), thermonuclear disruption of the disk is assured once $\alpha-$particles form.  This occurs because (1) the nuclear reactions that create (and destroy) deuterium (the building blocks of $\alpha-$particles) 
are still very fast compared to the evolution timescale of the disk (i.e., the disk is still in nuclear statistical equilibrium [NSE]) when the temperature drops sufficiently that the NSE $\alpha-$particle mass fraction becomes $\sim$ unity (e.g., Chen $\&$ Beloborodov 2007, eq.~[13]); and (2) the energy released by forming $\alpha-$particles ($\sim 7$ MeV per nucleon) exceeds the disk's gravitational binding energy (see MPQ08a,b).  Thus, the distributions in Figure $\ref{fig:yef}$ roughly bracket the distribution of $Y_{e}$ in the outflows that carry away the remaining mass in the disk at late times.

Figure $\ref{fig:yef}$ shows that $\approx 0.02-0.03M_{\sun}$ (or $\approx 20-30\%$ of the initial disk mass) remains when the disk is largely advective.  A comparable amount of mass remains when the disk forms $\alpha-$particles because this occurs at approximately the same time (MPQ08a,b; Beloborodov 2008).  Although the bulk of the disk is initially driven neutron-rich by degenerate electron captures, Figure $\ref{fig:yef}$ shows that a large portion of the mass remaining at late times has $Y_{e} \gtrsim 0.5$.  As discussed in $\S\ref{sec:bigpicture}$, the disk composition is driven proton-rich by the powerful $\nu_{e}$ flux from the central proto-NS.  

We have performed similar calculations for other disk parameters in order to study the sensitivity of our results to the disk's initial properties (see Table $\ref{table:yef}$).  For example, the disks found by D06,07 are fairly massive, but a less massive disk would result from a slower rotating WD.  Thus, we have also performed calculations using the same initial density profile from D06, but normalizing the total initial disk mass to $M_{d,0} = 0.01M_{\sun}$ instead of $0.1 M_{\sun}$.  We have also calculated models with $\Sigma(r,t=0)$, $L_{\nu}(t)$ and $\langle\epsilon_{\nu}\rangle(t)$ taken from the rapidly rotating, 1.92 $M_{\sun}$ WD model of D06, which formed a more massive initial disk ($M_{d,0} \approx 0.5M_{\sun}$).  In the high $M_{d,0}$ case we find that a significant fraction of the disk has exactly $Y_{e} = 0.5$ at freeze-out.  This occurs because in this model $\alpha-$particle formation precedes weak freeze-out; thus, any excess free neutrons or protons are almost immediately captured into $\alpha-$particles (which have $Y_{e} = 0.5$) following a neutrino absorption (the ``alpha-effect'' of Fuller $\&$ Meyer 1995).  

In Table \ref{table:yef} we summarize our results for the total mass ($M_{\rm adv}$) and the total mass with $0.485 \lesssim Y_{e} \lesssim 0.6$ ($M_{\rm Ni}$) when the disk has become $90\%$ advective.  The latter is labeled $M_{\rm Ni}$ because a significant fraction of the mass ejected with $Y_{e}$ in this range produces $^{56}$Ni (see $\S\ref{sec:miniSN}$).  In all cases, we find that a significant fraction ($\gtrsim 50\%$) of the mass remaining at late times has $Y_{e} \gtrsim 0.5$.  Thus, under a variety of conditions, AIC leads to $\sim 10\%$ of the initial disk being ejected as moderately proton-rich material.

\begin{table}
\begin{center}
\vspace{0.05 in}\caption{Late-Time Mass and Composition of AIC Disks}
\label{table:yef}

\begin{tabular}{lcccccc}
\hline
\hline
\multicolumn{1}{c}{$M_{d,0}^{(a)}$} &
\multicolumn{1}{c}{Model$^{(b)}$} &
\multicolumn{1}{c}{$\alpha$} &
\multicolumn{1}{c}{$M_{\rm adv}^{(c)}$} &
\multicolumn{1}{c}{$M_{\rm Ni}^{(d)}$} &
\\
($M_{\sun}$) & & & ($M_{\sun}$) & ($M_{\sun}$) \\
\hline 
\\
0.1 & $D06 (1.46M_{\sun}$WD) & 0.3 & 0.030 & 0.021 \\
0.1 & $D06 (1.46M_{\sun}$WD) & 0.03 & 0.021 & $8.9\times 10^{-3}$ \\
0.01 & $D06 (1.46M_{\sun}$WD) & 0.3 & $4.3\times 10^{-3}$ & $1.8\times 10^{-3}$ \\
0.5 & $D06 (1.92M_{\sun}$WD) & 0.3 & 0.13 & 0.066 \\
0.5 & $D06 (1.92M_{\sun}$WD) & 0.03 & 0.11 & 0.051 \\

\hline
\end{tabular}
\end{center}
{\small
(a) Initial disk mass; (b) Model used for the initial surface density $\Sigma(r,t=0)$ and the proto-NS neutrino luminosities and energies; (c) Mass remaining when the disk is $90\%$ advective; (d) Mass with $0.485 < Y_{e} < 0.6$ when the disk is $90\%$ advective.}
\end{table}
\vspace{-0.8 cm}
\section{Optical Transients from AIC}
\label{sec:miniSN}
\vspace{-0.2 cm}
In some ways, AIC can be considered a ``failed'' Type Ia SN because it does not produce $\sim 1M_{\sun}$ of shock-heated ejecta.  Indeed, the total $^{56}$Ni mass ejected from the proto-NS is $\lesssim 10^{-3}M_{\sun}$ (e.g., Woosley $\&$ Baron 1992; D06,07; Metzger, Thompson, $\&$ Quataert 2008), rendering isolated NS birth practically invisible.  In the presence of rotation, however, this failure is not complete.  In $\S\ref{sec:disk_evo}$ we showed that if an accretion disk forms around the proto-NS during collapse, a natural consequence of the disk's subsequent evolution is that up to $\sim 10^{-2}M_{\sun}$ of material with $Y_{e} \gtrsim 0.5$ is ejected within a few seconds after core bounce.

When hyper-accreting disks become radiatively inefficient and begin forming $\alpha-$particles, gas and radiation pressure in the midplane are comparable and the entropy is $S \sim 10$ k$_{B}$ baryon$^{-1}$.  Subsequent outflows from the disk are likely to possess a comparable entropy because heating that occurs as matter is unbound (e.g., due to $\alpha-$particle formation) is unlikely to deposit more than a few k$_{B}$ baryon$^{-1}$.  A typical outflow speed is $v \sim 0.1-0.2$ c: this is approximately the escape speed from the disk when it becomes advective and a similar speed is attained from the $\approx 9$ MeV baryon$^{-1}$ released in producing Fe-peak elements.  

\begin{figure}
\resizebox{\hsize}{!}{\includegraphics{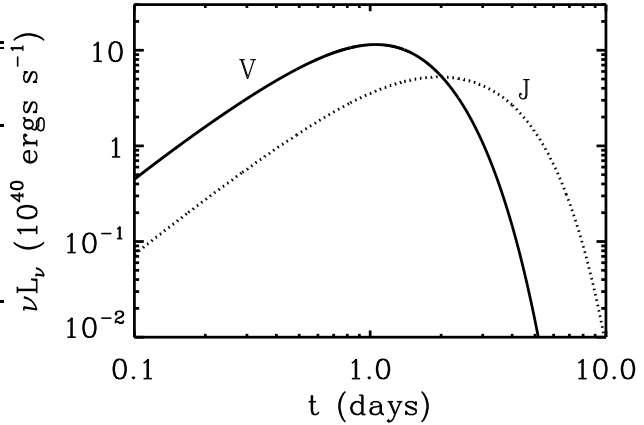}}
\caption{Luminosity of Ni decay-powered transient as a function of time for Ni mass $M_{\rm Ni} = 0.01M_{\sun}$, total ejected mass $M_{\rm tot} = 0.02M_{\sun}$, and ejecta velocity $v = 0.1$ c.  The luminosities in V and J-Band (0.44 and 1.26 $\mu$m, respectively) are shown with a solid and dashed line, respectively.  }
\label{fig:lightcurve}
\end{figure}

Matter with $0.485 \lesssim Y_{e} \lesssim 0.6$ that begins in NSE at low entropy and expands adiabatically at mildly relativistic speeds is primarily synthesized into $^{56}$Ni (e.g., Pruet et al.~2004) because $^{56}$Ni is favored in NSE under proton-rich conditions (Seitenzahl et al.~2008).  Our results in Figure \ref{fig:yef} and Table \ref{table:yef} therefore imply that AIC outflows yield a typical Ni mass $M_{\rm Ni} \sim 10^{-2}M_{\sun}$ if the rotating 1.46$M_{\sun}$-WD progenitors used in the calculations of D06,07 are representative.  The decay of this Ni can reheat the (adiabatically cooled) ejecta sufficiently to produce detectable transient emission once the outflow expands sufficiently that photons can diffuse out (Arnett 1982).    

We explore this possibility by calculating the light curves of ejecta heated by Ni decay using the method of Kulkarni (2005) and MPQ08a.  Figure \ref{fig:lightcurve} shows the V and J-band luminosities as a function of time since collapse for an outflow with Ni mass $M_{\rm Ni} = 10^{-2}M_{\sun}$ and total ejecta mass $M_{\rm tot} = 2\times 10^{-2}M_{\sun}$ expanding at $v = 0.1$ c.  The total luminosity peaks in the visual (V-band) on a $\sim 1$ day timescale, with $\nu L_{\nu} \sim 10^{41}$ ergs s$^{-1}$.  The J-band light curve peaks somewhat later than V-band because the temperature at the photosphere decreases as the material expands.  We discuss the likelihood of detecting such a transient with blind surveys in $\S\ref{sec:rates}$. 
   
Although $^{56}$Ni is likely to be the dominant isotope synthesized, other isotopes will also be produced in smaller abundances.  For instance, under the conditions of interest $\sim 10\%$ of the ejected mass may remain in the form of $\alpha-$particles (Woosley $\&$ Hoffman 1992).  Furthermore, our low-$\alpha$ calculations in Figure \ref{fig:yef} indicate that some fraction of the material ejected may have $Y_{e} \lesssim 0.485$ (Fig.~\ref{fig:yef}).  In this case, freeze-out from NSE produces both stable (e.g., $^{58,60,62,64}$Ni, $^{54,56,58}$Fe) and radioactive (e.g., $^{66,68,70}$Ni, $^{60,62,66}$Zn) Fe-group elements instead of $^{56}$Ni (Woosley $\&$ Hoffman 1992).  Although this material will contribute to the opacity of the outflow ($M_{\rm tot}$ above), its additional contribution to the radioactive heating is unlikely to substantially alter the purely Ni-powered light curves in Figure \ref{fig:lightcurve}.

\section{Detection Prospects}
\label{sec:rates}
\vspace{-0.2 cm}
Because AIC has never been observationally identified, its rate is uncertain.  Binary population synthesis models predict galactic rates in the range $R_{\rm AIC} \sim 10^{-6}-10^{-4}$ yr$^{-1}$ (e.g., Yungelson $\&$ Livio 1998).  The amount of highly neutron-rich material ejected per AIC can also be used to constrain the AIC rate (Hartmann et al.~1985; Fryer et al.~1999): if the $\sim 3\times 10^{-3}M_{\sun}$ of ejected material with $Y_{e} \lesssim 0.4$ in the calculations of D06 is representative, $R_{\rm AIC}$ must be $\lesssim 10^{-4}$ yr$^{-1}$ so as not to over-produce the abundances of neutron-rich isotopes in our solar system.  If the proto-NS forms with a strong large-scale magnetic field $\sim 10^{15}$ G, then a much larger quantity $\sim 0.1 M_{\sun}$ of neutron-rich material is ejected by the proto-NS's magneto-centrifugally driven wind (D07; Metzger, Thompson, $\&$ Quataert 2008) and the AIC rate is constrained to be even lower.

Assuming that the AIC rate is proportional to the blue stellar luminosity (Phinney 1991), a Galactic rate of $R_{\rm AIC} \equiv 10^{-4}R_{-4}$ yr$^{-1}$ corresponds to a volumetric rate of $10^{-6}R_{-4}$ Mpc$^{-3}$ yr$^{-1}$.  For the 1.46$M_{\sun}$-WD model of D06, we predict that AIC produces an optical transient with a peak luminosity $\sim 10^{41}$ ergs s$^{-1}$ (Fig.~\ref{fig:lightcurve}).  This corresponds to a maximum detection distance of 650(410)[100] Mpc for a limiting magnitude of 25(24)[21].  Thus, the PanSTARRs Medium Deep Survey (MDS), which covers $\sim 84$ deg$^{2}$ in $g$ and $r$ down to AB magnitude 25, will detect $\sim 2.4 R_{-4}$ yr$^{-1}$.  The Palomar Transient Factory (PTF), which surveys $\sim 8000$ deg$^{2}$ to a limiting AB magnitude of 21, will detect $0.9R_{-4}$ yr$^{-1}$.  Thus, if $R_{\rm AIC} \sim 10^{-4}$ yr$^{-1}$, MDS and PTF should detect a few AIC events per year.  Prospects for detection are much better with the Large Synoptic Survey Telescope (LSST), which will image the entire sky down to a limiting magnitude $\sim 24.5$ every 3-4 nights and should detect AIC events at a rate $\sim 800R_{-4}$ yr$^{-1}$.  The short, few-day timescale of the transient requires observations with high cadence.  Shallow, wider-area surveys such as PTF may be preferable because separating the transient from its host galaxy's light may be difficult for high redshift events.

The detectability of an optical transient is sensitive to whether there is sufficient mass in a remnant disk, which in turn depends on the angular momentum distribution of the WD progenitor.  The very massive AIC progenitor from D06 (1.92 $M_{\sun}$) is consistent with the large shear found by Yoon $\&$ Langer (2004, 2005), who studied angular momentum transport by Kelvin-Helmholtz instabilities during accretion.  Piro (2008) argued, however, that large magnetic stresses impose nearly solid body rotation, which prevents the WD from reach a mass much over the (non-rotating) Chandrasekhar limit (this may also be the case for purely hydrodynamic instabilities; see Saio $\&$ Nomoto 2004).  More work is needed to understand the possible progenitors of AIC, and the resulting implications for the accretion disk formed during collapse, taking into account the effects of magnetic stresses and compositional gradients on the redistribution of angular momentum during accretion.

Other thermal transients are predicted to occur in nature on $\sim$ day timescales, such as ``.Ia'' SNe due to unstable thermonuclear He flashes from WD binaries (Bildsten et al.~2007).  Although such events may originate from a similar stellar population and could be confused with AIC, a defining characteristic of AIC outflows is that most of the ejected material is processed through low entropy NSE.  As a result, we expect few intermediate-mass elements to be produced (such as O, Ca, and Mg), although small amounts of He may be present.  Thus, the Ni-rich explosions produced by AIC should be spectroscopically distinct from other types of SN identified to date.  

The timescale, velocity, and composition of the AIC transient can be modified if significant amounts of WD material remain at large radii following AIC, either because the WD is still collapsing at late times or because some matter remains centrifugally supported in a remnant disk created by a super-M$_{\rm Ch}$ WD-WD merger (e.g., Yoon et al. 2007).  In the later case, depending on the mass and composition of the WD binary, up to $\sim 0.1 M_{\sun}$ in C, O, Ne, or He could remain at a radius $R \sim 10^{9}$ cm.  Once the Ni wind (with energy $\sim 10^{50}$ ergs) impacts this remnant material, it will (1) shock heat the material to a few times $10^{9}$ K, which may synthesize intermediate mass elements (Woosley et al.~2002), but will not disassociate the Ni and will likely leave some unburned WD material; (2) slow the ejecta from $v \sim 0.1$ c to a few thousand km s$^{-1}$.  A combination of slower ejecta and higher opacity would also lengthen the duration of the light curve by a factor of $\sim 10$ from that shown in Figure $\ref{fig:lightcurve}$, making such an event easier to detect.  Indeed, such an ``enshrouded AIC'' model represents a possible explanation for sub-luminous, sub-Chandrasekhar Type I SNe, such as 2008ha, which rose to its peak brightness in only $\sim 10$ days, possessed low line velocities ($\sim 2000$ km s$^{-1}$), and was inferred to have $M_{\rm Ni} \approx 3\times 10^{-3}M_{\sun}$ and $M_{\rm tot} \sim 0.15M_{\sun}$ (Foley et al.~2009).  We plan to study this possibility further, as well as explore the interaction between the disk wind and the outoing SN shock, in future work.

AIC may also be detectable with upcoming km-scale gravitational wave (GW) detectors such as Advanced LIGO and VIRGO.  If the collapse remains axisymmetric, the detection of AIC in GWs appears unlikely except within the Milky Way (D06,07; Dimmelmeier et al.~2008).  However, if the progenitor WD is very rapidly rotating (as may result following a WD-WD merger), the ratio of rotational to gravitational energy of the proto-NS upon collapse may be sufficiently large for the growth of nonaxisymmetric instabilities, which would greatly increase the GW signal and detection prospects (Fryer et al.~2002; Ott 2008).  Indeed, D06 find that their their 1.92 $M_{\sun}$-WD model rotates sufficiently rapidly to trigger a dynamically unstable spiral mode upon collapse (Ott et al.~2005).  We note that our calculations predict that large Ni yields will accompany the massive disks that form from such rapidly rotating progenitors (e.g., we find $M_{\rm Ni} \approx 0.07M_{\sun}$ for D06's 1.92$M_{\sun}$ model; see Table $\ref{table:yef}$).  Thus, events accompanied by GW emission that is detectable to extragalactic distances will likely produce a somewhat brighter SN transient than our fiducial example in Figure $\ref{fig:lightcurve}$.  More generally, the brightness of the optical emission from AIC directly traces the importance of rotation during the collapse.
\vspace{-0.8 cm}

\section*{Acknowledgments}
\vspace{-0.2 cm}
We thank Luc Dessart for providing his AIC calculations.  We also thank Todd Thompson for helpful conversations and for providing the neutrino luminosity and spectral evolution from his rotating core-collapse supernova calculations.  A.~L.~P.~is supported by the Theoretical Astrophysics Center at UC Berkeley.  B.~D.~M and E.~Q.~were supported in part by the David and Lucile Packard Foundation, NASA grant NNG05GO22H, and the NSF-DOE Grant PHY-0812811.

\label{lastpage}


\vspace{-0.8 cm}
\begin{thebibliography}{99}

\bibitem[Arnett(1982)]{1982ApJ...253..785A} Arnett, W.~D.\ 1982, \apj, 253, 
785

\bibitem[Bailyn 
\& Grindlay(1990)]{1990ApJ...353..159B} Bailyn, C.~D., \& Grindlay, J.~E.\ 1990, \apj, 353, 159 

\bibitem[Beloborodov(2008)]{2008AIPC.1054...51B} Beloborodov, A.~M.\ 2008, 
AIP Conference Series, 1054, 51 

\bibitem[Bildsten et al.(2007)]{2007ApJ...662L..95B} Bildsten, L., Shen, 
K.~J., Weinberg, N.~N., \& Nelemans, G.\ 2007, \apjl, 662, L95 

\bibitem[Blandford \& Begelman(1999)]{1999MNRAS.303L...1B} Blandford, R.~D., \& Begelman, M.~C.\ 1999, \mnras, 303, L1 

\bibitem[Brandenburg, A.]{Brandenburg, A.} Brandenburg, A. 2003, in Advances in Nonlinear Dynamos: The Fluid Mechanics of Astrophysics \& Geophysics, Vol 9, ed. A. Ferriz-Mas \& M. Nunez, 269

\bibitem[Burrows \& Lattimer(1986)]{1986ApJ...307..178B} Burrows, A., \& 
Lattimer, J.~M.\ 1986, \apj, 307, 178 

\bibitem[Burrows 
\& Mazurek(1983)]{1983Natur.301..315B} Burrows, A., \& Mazurek, T.~L.\ 1983, \nat, 301, 315 

\bibitem[Canal et 
al.(1990)]{1990ARA&A..28..183C} Canal, R., Isern, J., \& Labay, J.\ 1990, ARAA, 28, 183 

\bibitem[Chen 
\& Beloborodov(2007)]{2007ApJ...657..383C} Chen, W.-X., \& Beloborodov, A.~M.\ 2007, \apj, 657, 383 

\bibitem[Dessart et al.(2006)]{2006ApJ...644.1063D} Dessart, L., Burrows, 
A., Ott, C.~D., Livne, E., Yoon, S.-C., 
\& Langer, N.\ 2006, \apj, 644, 1063 $(D06$)

\bibitem[Dessart et al.(2007)]{2007ApJ...669..585D} Dessart, L., Burrows, 
A., Livne, E., \& Ott, C.~D.\ 2007, \apj, 669, 585 $(D07)$

\bibitem[DiMatteo et al.(2002)]{dim02}
DiMatteo, T., Perna, R., \& Narayan, R. 2002, \apj, 579, 706 

\bibitem[Dimmelmeier et al.(2008)]{2008PhRvD..78f4056D} Dimmelmeier, H., 
Ott, C.~D., Marek, A., \& Janka, H.-T.\ 2008, Phys.~Rev.~D, 78, 064056 

\bibitem[Foley et al.(2009)]{2009arXiv0902.2794F} Foley, R.~J., et al.\ 
2009, arXiv:0902.2794 

\bibitem[Fryer et al.(1999)]{1999ApJ...516..892F} Fryer, C., Benz, W., 
Herant, M., \& Colgate, S.~A.\ 1999, \apj, 516, 892 

\bibitem[Fryer et al.(2002)]{2002ApJ...565..430F} Fryer, C.~L., Holz, 
D.~E., \& Hughes, S.~A.\ 2002, \apj, 565, 430 

\bibitem[Fuller 
\& Meyer(1995)]{1995ApJ...453..792F} Fuller, G.~M., \& Meyer, B.~S.\ 1995, \apj, 453, 792 

\bibitem[Grindlay(1987)]{1987IAUS..125..173G} Grindlay, J.~E.\ 1987, The 
Origin and Evolution of Neutron Stars, 125, 173 

\bibitem[Grindlay 
\& Bailyn(1988)]{1988Natur.336...48G} Grindlay, J.~E., \& Bailyn, C.~D.\ 1988, \nat, 336, 48 

\bibitem[Guti{\'e}rrez et 
al.(2005)]{2005A&A...435..231G} Guti{\'e}rrez, J., Canal, R., \& Garc{\'{\i}}a-Berro, E.\ 2005, \aap, 435, 231 

\bibitem[Hartmann et al.(1985)]{1985ApJ...297..837H} Hartmann, D., Woosley, 
S.~E., \& El Eid, M.~F.\ 1985, \apj, 297, 837

\bibitem[Kulkarni(2005)]{2005astro.ph.10256K} Kulkarni, S.~R.\ 2005, ArXiv 
Astrophysics e-prints, arXiv:astro-ph/0510256 

\bibitem[Lee \& Ramirez-Ruiz(2007)]{lr07}
Lee, W. \& Ramirez-Ruiz, E. 2007, New J. Phys., 9, 17 

\bibitem[Metzger et al.(2008)]{2008MNRAS.tmp.1055M} Metzger, B.~D., Piro, 
A.~L., \& Quataert, E.\ 2008, \mnras, 390, 781 $(MPQ08a)$

\bibitem[Metzger et al.(2008)]{2008arXiv0810.2535M} Metzger, B.~D., Piro, 
A.~L., \& Quataert, E.\ 2008, Accepted to MNRAS, arXiv:0810.2535 $(MPQ08b)$

\bibitem[Metzger et al.(2008)]{2008ApJ...676.1130M} Metzger, B.~D., 
Thompson, T.~A., \& Quataert, E.\ 2008, \apj, 676, 1130 

\bibitem[Michel(1987)]{1987Natur.329..310M} Michel, F.~C.\ 1987, \nat, 329, 
310

\bibitem[Miyaji et al.(1980)]{1980PASJ...32..303M} Miyaji, S., Nomoto, K., 
Yokoi, K., \& Sugimoto, D.\ 1980, PASJ, 32, 303 

\bibitem[Mochkovitch 
\& Livio(1989)]{1989A&A...209..111M} Mochkovitch, R., \& Livio, M.\ 1989, \aap, 209, 111 

\bibitem[Mochkovitch 
\& Livio(1990)]{1990A&A...236..378M} Mochkovitch, R., \& Livio, M.\ 1990, \aap, 236, 378 

\bibitem[Napiwotzki et 
al.(2002)]{2002A&A...386..957N} Napiwotzki, R., et al.\ 2002, \aap, 386, 957 

\bibitem[Napiwotzki et al.(2004)]{2004ASPC..318..402N} Napiwotzki, R., et 
al.\ 2004, Spectroscopically and Spatially Resolving the Components of the 
Close Binary Stars, 318, 402 

\bibitem[Nomoto \& Kondo(1991)]{1991ApJ...367L..19N} Nomoto, K., \& Kondo, 
Y.\ 1991, \apjl, 367, L19 

\bibitem[Ott et al.(2005)]{2005ApJ...625L.119O} Ott, C.~D., Ou, S., 
Tohline, J.~E., \& Burrows, A.\ 2005, \apjl, 625, L119 

\bibitem[Ott(2008)]{2008arXiv0809.0695O} Ott, C.~D.\ 2008, arXiv:0809.0695 

\bibitem[Phinney(1991)]{1991ApJ...380L..17P} Phinney, E.~S.\ 1991, \apjl, 
380, L17 

\bibitem[Piro(2008)]{2008ApJ...679..616P} Piro, A.~L.\ 2008, \apj, 679, 616 

\bibitem[Poelarends et al.(2008)]{2008ApJ...675..614P} Poelarends, 
A.~J.~T., Herwig, F., Langer, N., \& Heger, A.\ 2008, \apj, 675, 614 

\bibitem[Pons et al.(1999)]{1999ApJ...513..780P} Pons, J.~A., Reddy, S., 
Prakash, M., Lattimer, J.~M., \& Miralles, J.~A.\ 1999, \apj, 513, 780 

\bibitem[Pruet et al.(2004)]{2004ApJ...606.1006P} Pruet, J., Thompson, 
T.~A., \& Hoffman, R.~D.\ 2004, \apj, 606, 1006 

\bibitem[Pruet et al.(2003)]{2003ApJ...586.1254P} Pruet, J., Woosley, 
S.~E., \& Hoffman, R.~D.\ 2003, \apj, 586, 1254 

\bibitem[Qian et al.(1993)]{1993PhRvL..71.1965Q} Qian, Y.-Z., Fuller, 
G.~M., Mathews, G.~J., Mayle, R.~W., Wilson, J.~R., 
\& Woosley, S.~E.\ 1993, Physical Review Letters, 71, 1965 

\bibitem[] Saio, H., \& Nomoto, K.\ 2004, \apj, 615, 444

\bibitem[Seitenzahl et al.(2008)]{2008arXiv0808.2033S} Seitenzahl, I.~R., 
Timmes, F.~X., Marin-Lafl{\`e}che, A., Brown, E., Magkotsios, G., 
\& Truran, J.\ 2008, ArXiv e-prints, 808, arXiv:0808.2033 

\bibitem[Shakura \& Sunyaev(1973)]{ss73}
Shakura, N. I., \& Sunyaev, R. A. 1973, A\&A, 24, 337 

\bibitem[Shapiro 
\& Lightman(1976)]{1976ApJ...207..263S} Shapiro, S.~T., \& Lightman, A.~P.\ 1976, \apj, 207, 263 

\bibitem[Thompson et al.(2005)]{2005ApJ...620..861T} Thompson, T.~A., 
Quataert, E., \& Burrows, A.\ 2005, \apj, 620, 861 

\bibitem[van den Heuvel(1984)]{1984JApA....5..209V} van den Heuvel, 
E.~P.~J.\ 1984, Journal of Astrophysics and Astronomy, 5, 209 

\bibitem[Woosley \& Baron(1992)]{1992ApJ...391..228W}
Woosley, S. E. \& Baron, E. 1992, \apj, 391, 228

\bibitem[Woosley 
\& Hoffman(1992)]{1992ApJ...395..202W} Woosley, S.~E., \& Hoffman, R.~D.\ 1992, \apj, 395, 202 

\bibitem[Woosley et al.(2002)]{2002RvMP...74.1015W} Woosley, S.~E., Heger, 
A., \& Weaver, T.~A.\ 2002, Reviews of Modern Physics, 74, 1015 

\bibitem[]{} Yoon, S.-C., \& Langer, N.\ 2004, \aap, 419, 623

\bibitem[]{} Yoon, S.-C., \& Langer, N.\ 2005, \aap, 435, 967

\bibitem[Yoon et al.(2007)]{2007MNRAS.380..933Y} Yoon, S.-C., 
Podsiadlowski, P., \& Rosswog, S.\ 2007, \mnras, 380, 933 

\bibitem[Yungelson 
\& Livio(1998)]{1998ApJ...497..168Y} Yungelson, L., \& Livio, M.\ 1998, \apj, 497, 168


%
%
%
%
%

\end{thebibliography}
\end{document}